\documentclass[final,5p,times,twocolumn]{elsarticle}
\usepackage{amssymb}
\usepackage{amsmath}
\usepackage{lipsum}

\usepackage{lineno}
\usepackage{xcolor}

\journal{Journal of Alloys and Compounds}

\begin{document}

\begin{frontmatter}

\title{Microscopic Investigation of the Superconducting State in CuCo$_{2}$S$_{4}$: Evidence for an Intermediate-Coupling Fully Gapped Superconductor}

\author[first]{K. Panda}
\affiliation[first]{organization={Department of Physics, Ariel University},
            city={Ariel},
            postcode={40700},
            country={Israel}}

\author[second]{A. Bhattacharyya}
\affiliation[second]{organization={Department of Physics, Ramakrishna Mission Vivekananda Educational and Research Institute},
            city={Belur Math, Howrah},
            postcode={711202},
            state={West Bengal},
            country={India}}

\author[third]{Liang-Wen Ji}
\affiliation[third]{organization={School of Physics, Interdisciplinary Center for Quantum Information, and State Key Lab of Silicon and Advanced Semiconductor Materials, Zhejiang University},
            city={Hangzhou},
            postcode={310058},
            country={China}}

\author[third]{Jing Li}

\author[fourth]{R. Stewart}
\affiliation[fourth]{organization={ISIS Neutron and Muon Source, Rutherford Appleton Laboratory},
            city={Chilton, Didcot},
            postcode={OX11 0QX},
            state={Oxon},
            country={United Kingdom}}

\author[fourth,fifth]{D. T. Adroja}
\affiliation[fifth]{organization={Highly Correlated Matter Research Group, Physics Department, University of Johannesburg},
            city={Auckland Park},
            postcode={2006},
            country={South Africa}}

\author[third,sixth]{Guang-Han Cao}
\affiliation[sixth]{organization={Collaborative Innovation Centre of Advanced Microstructures, Nanjing University},
            city={Nanjing},
            postcode={210093},
            country={China}}

\begin{abstract}
\textcolor{black}{The thiospinel compound CuCo$_2$S$_4$ provides an attractive platform for exploring superconductivity in transition-metal chalcogenide spinels. Here, we report the first microscopic investigation of the superconducting state in CuCo$_2$S$_4$ using muon spin rotation and relaxation ($\mu$SR) measurements, complemented by magnetization and heat-capacity experiments. The temperature dependence of the superconducting depolarization rate obtained from transverse-field $\mu$SR measurements indicates a fully gapped superconducting order parameter. The extracted gap ratio $2\Delta(0)/(k_{\mathrm{B}}T_\mathrm{SC}) = 3.95(2)$ exceeds the BCS weak-coupling value of 3.53, placing CuCo$_2$S$_4$ in the intermediate electron-phonon coupling regime. Zero-field $\mu$SR measurements were performed to probe possible time-reversal symmetry breaking (TRSB) in the superconducting state. Within the experimental resolution, no additional spontaneous internal magnetic fields are observed below $T_c$. However, due to the presence of a ferromagnetic impurity phase and the associated fast-relaxing signal component, the sensitivity of the present measurements to weak spontaneous fields is reduced. Consequently, while no evidence for TRSB is detected, its existence cannot be definitively ruled out. Overall, our combined thermodynamic and $\mu$SR results demonstrate that CuCo$_2$S$_4$ exhibits a fully gapped superconducting state with intermediate coupling strength, consistent with conventional $s$-wave superconductivity in this cobalt-based thiospinel system.}
\end{abstract}


\end{frontmatter}

\section{Introduction}
\label{introduction}

Thiospinel compounds, a subset of the spinel family, are distinguished by their crystal structures and chemical compositions~\cite{Naumann2008,Bosi2019}. The spinel structure follows the general formula AB$_2$X$_4$, where A and B are metal cations and X is an anion such as oxygen (O), sulfur (S), or selenium (Se). While superconductivity in spinels is exemplified by LiTi$_2$O$_4$, which undergoes a superconducting transition at 14\,K~\cite{Johnston1976,McCallum1976}, thiospinels, where sulfur replaces oxygen, exhibit distinct electronic, magnetic, and structural properties. Superconductivity in thiospinels is rare but has been observed in compounds such as CuCo$_2$S$_4$, CuRh$_2$S$_4$, and Cu$_{1-x}$Zn$_x$Ir$_2$S$_4$ (maximum $T_\mathrm{SC} = 3.4$\,K for $x=0.3$)~\cite{Naumann2008,Bosi2019}. These materials also offer the possibility of coexisting superconductivity and magnetic ordering, positioning thiospinels as model systems for studying the interplay between superconductivity and magnetism~\cite{Naumann2008,Bosi2019}.

Superconductivity in CuRh$_2$S$_4$ was first reported in 1967, with a transition temperature ranging from 4.35 to 4.8\,K~\cite{VanMaaren1967,Robbins1967,Hagino1995}. Similarly, CuRh$_2$Se$_4$ exhibits a superconducting transition at 3.48\,K~\cite{Robbins1967}. Specific heat measurements for CuRh$_2$S$_4$ reveal a normalized heat capacity jump at $T_\mathrm{SC}$ of 1.82~\cite{Hagino1995}, while NMR studies point to electron-phonon interactions as the dominant mechanism, classifying these compounds as weak- to intermediate-coupling superconductors~\cite{Kumagai1995}. In contrast, CuIr$_2$S$_4$ does not exhibit intrinsic superconductivity; however, superconductivity can be induced through Zn for Cu substitution, with transition temperatures reaching up to 3.4\,K~\cite{Cao2001,Suzuki1999}. These observations demonstrate that superconductivity in thiospinels is highly sensitive to stoichiometry and electronic structure modifications~\cite{VanMaaren1967,Robbins1967}.

Previous studies on Cu$_{1+x}$Co$_{2-x}$S$_4$ (with $x$ ranging from 0 to 0.5) reported Curie-Weiss-type susceptibility, indicative of antiferromagnetic behavior below 19\,K, accompanied by the emergence of superconductivity with an onset transition temperature, $T_\mathrm{SC}^{\text{onset}}$, at approximately 4.0\,K~\cite{Aito2004}. In contrast, no evidence of an antiferromagnetic phase was observed in CuCo$_2$S$_4$ in samples reported by G.~H.~Cao and collaborators~\cite{Jin2021}. Furthermore, sulfur-deficient CuCo$_2$S$_{3.7}$ exhibited no superconductivity down to 1.8\,K, underscoring the critical role of complete sulfurization for the emergence of superconductivity~\cite{Jin2021}. Natural carrollite samples, such as Cu$_{0.92}$Co$_{2.07}$S$_4$, have demonstrated superconductivity around 3\,K~\cite{Gibson2023}. Band structure calculations reveal that the electronic structure of CuCo$_2$S$_4$ is predominantly governed by Co $3d$ and S $3p$ states, with significant hybridization between these orbitals~\cite{Oda1995,Yue2021}. Recent theoretical studies suggest that electron-phonon coupling drives superconductivity in CuCo$_2$S$_4$, despite the presence of partially magnetic Co $3d$ states~\cite{Skornyakov2023,HemmatiEslamlu2023,Habibi2023}.

While cobalt is typically ferromagnetic in its elemental form, there are a few known examples of cobalt-based superconductors. For instance, the non-centrosymmetric superconductor ThCoC$_2$ displays a nodal superconducting gap structure while preserving time-reversal symmetry, suggesting unconventional pairing mechanisms inconsistent with standard BCS theory~\cite{BhattacharyyaThCoC2}. In contrast, the centrosymmetric superconductor Sc$_5$Co$_4$Si$_{10}$ exhibits an isotropic, fully gapped superconducting state accompanied by evidence of time-reversal symmetry breaking~\cite{Bhattacharyya2022}. In CuCo$_2$S$_4$, the lack of magnetic ordering despite Co's magnetic moment underlines the dominance of electron-phonon coupling over magnetic interactions. In CuCo$_2$S$_4$, studies on the symmetry of the superconducting (SC) gap have yielded conflicting conclusions, reflecting the system's complexity. Early nuclear magnetic resonance (NMR) investigations suggested a gapless SC state, attributing this phenomenon to enhanced antiferromagnetic fluctuations~\cite{Furukawa1995}. However, subsequent NMR studies indicated a fully isotropic SC gap with a gap-to-$T_\mathrm{SC}$ ratio of $2\Delta/k_\mathrm{B}T_\mathrm{SC} = 4.14$, along with weak antiferromagnetic spin correlations~\cite{Sugita2000,Wada2001}. Recent specific heat measurements further confirm the absence of a gapless or nodal SC state in nearly stoichiometric CuCo$_2$S$_4$ samples~\cite{Jin2021}. Interestingly, related cobalt-based systems, such as Na$_x$CoO$_2 \cdot y$H$_2$O, have been associated with an anisotropic order parameter~\cite{Baskaran2003,Fujimoto2004,Wang2004}. The interplay between superconductivity and magnetism in thiospinels makes them promising candidates for studying competition or coexistence between these orders.

To date, no microscopic investigations of the superconducting properties of CuCo$_2$S$_4$ have been reported. In this study, we present a detailed analysis of its superconducting behavior using \textcolor{black}{muon spin rotation and relaxation} ($\mu$SR) measurements. Our results confirm that CuCo$_2$S$_4$ exhibits a fully gapped superconducting order parameter with a gap-to-critical temperature ratio, $2\Delta/k_\mathrm{B}T_\mathrm{SC} = 3.95(2)$, consistent with previous NMR studies and exceeding the BCS weak-coupling limit of 3.53. While zero-field $\mu$SR measurements show no definitive evidence of spontaneous internal magnetic fields in the superconducting state, the presence of a minor ferromagnetic impurity phase precludes a conclusive assessment of time-reversal symmetry breaking.

\begin{figure*}[t]
\centering
\includegraphics[width=0.9\linewidth]{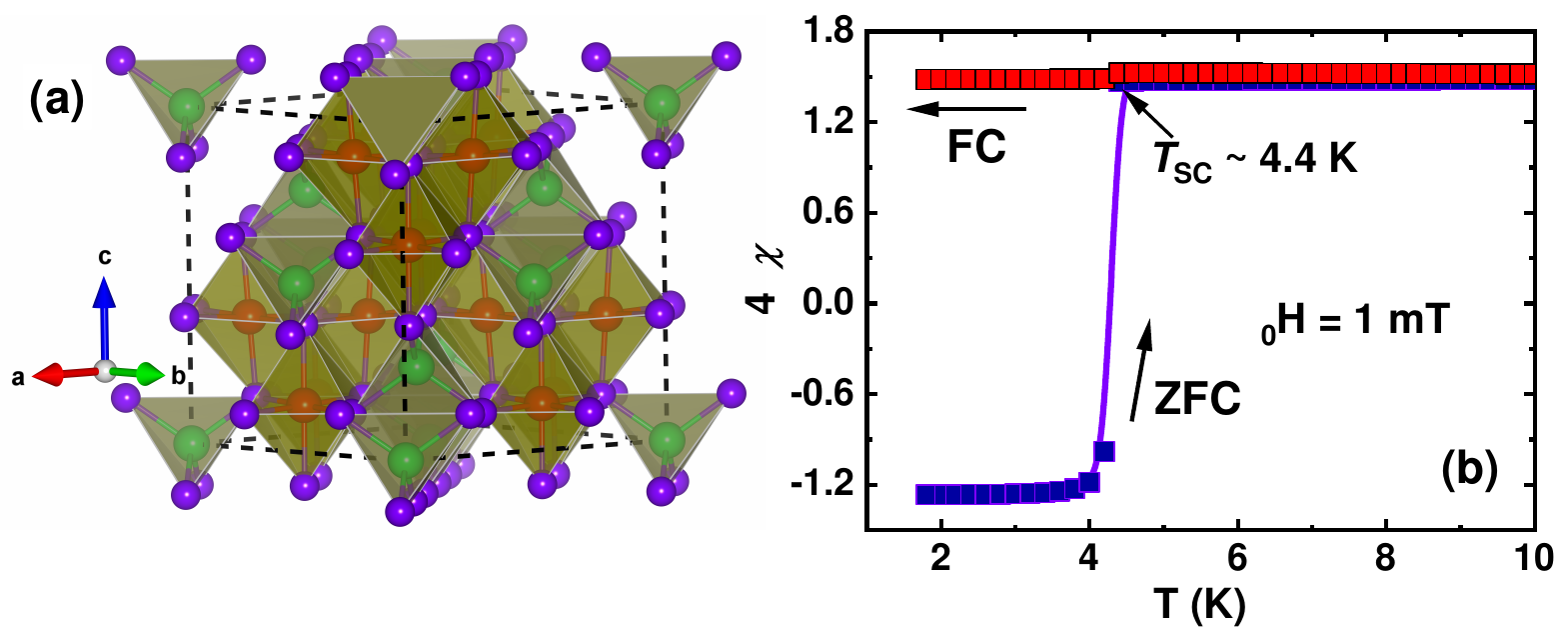}
\caption{(a) Crystal structure of CuCo$_{2}$S$_{4}$, illustrating the arrangement of tetrahedral (CuS$_{4}$) and octahedral (CoS$_{6}$) units. In this depiction, copper (Cu) atoms are represented as green spheres (large), cobalt (Co) atoms as red spheres (medium), and sulfur (S) atoms as violet spheres (small). (b) Temperature-dependent magnetic susceptibility measured under zero-field-cooled (ZFC) and field-cooled (FC) conditions in an applied magnetic field of 1\,mT.}
\label{fig1}
\end{figure*}

\begin{figure}[t]
\centering
\includegraphics[width=0.9\linewidth]{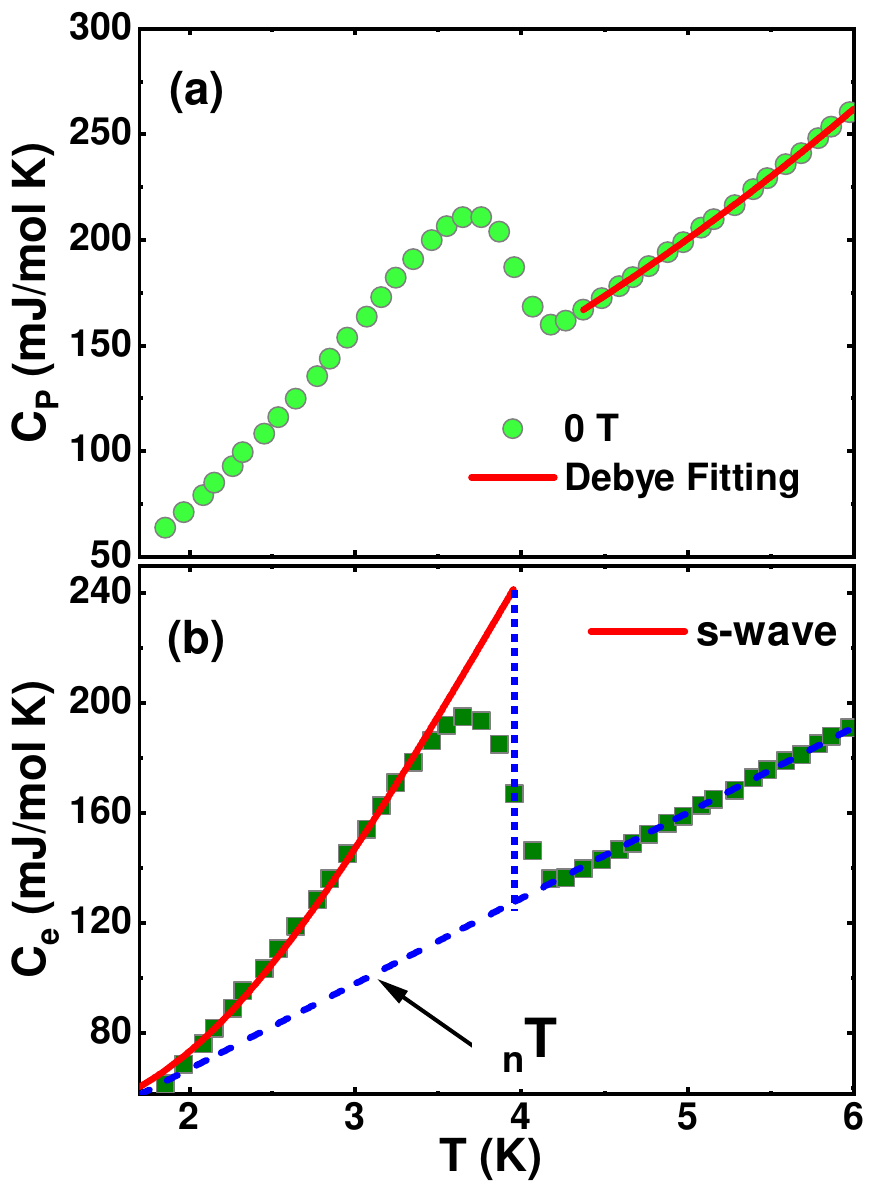}
\caption{(a) Temperature dependence of the heat capacity, \( C_\mathrm{P}(T) \), measured from 1.8\,K to 6\,K in zero applied magnetic field. The solid line represents a fit of \( C_\mathrm{P}(T) \) in the normal state using the Debye model. (b) Electronic contribution to the zero-field heat capacity, \( C_\mathrm{e} \), as a function of temperature. The solid line shows a fit to \( C_\mathrm{e} \) using the isotropic Bardeen-Cooper-Schrieffer (BCS) model, including a term for the nonsuperconducting fraction, \( \gamma_\mathrm{ns}T \). The blue dashed line represents an extrapolation of the normal-state electronic heat capacity, \( \gamma_\mathrm{n}T \).}
\label{fig2}
\end{figure}

\section{Experimental Techniques}
\label{experimental}
\textcolor{black}{Polycrystalline samples of CuCo$_{2}$S$_{4}$ were synthesized using a two-step solid-state reaction method. High-purity elemental powders of Cu (99.997\%), Co (99.998\%), and S (99.999\%) were used as starting materials. The powders were weighed according to the desired stoichiometry, thoroughly mixed in an agate mortar under an inert atmosphere, and sealed in an evacuated quartz ampoule to prevent oxidation and sulfur loss during high-temperature treatment. In the first stage, the precursor compound CuCo$_{2}$S$_{3.7}$ was prepared by mixing the elemental powders in the nominal composition. The mixture was sealed in an evacuated quartz ampoule and annealed at 750$^\circ$ C for 72 h. After the initial reaction, the sample was reground, pelletized, and subjected to a second annealing step under the same conditions to improve homogeneity and phase formation. In the second stage, the sulfur-deficient precursor was sulfurized to obtain the stoichiometric CuCo$_{2}$S$_{4}$ phase. This was achieved by sealing the precursor together with additional elemental sulfur (approximately 0.35 S per formula unit) in an evacuated silica ampoule and annealing the mixture at 450$^\circ$for 144 h. This sulfur-compensation step allows the restoration of the correct sulfur stoichiometry and promotes the formation of the superconducting CuCo$_{2}$S$_{4}$ phase. The amount of sulfur added during the sulfurization step was carefully controlled by weighing the elemental sulfur using a high-precision analytical balance prior to sealing the ampoule. The sealed-ampoule environment ensures a controlled sulfur vapor pressure during annealing, which minimizes sulfur loss and helps maintain the intended stoichiometry. We found that the superconducting properties of the final compound are sensitive to the sulfur content, and therefore the sulfurization conditions were optimized to obtain the highest phase purity and reproducible superconducting behavior.}

 \textcolor{black}{The XRD measurements were performed using a PANalytical Empyrean Series 2 diffractometer with monochromatic Cu-K$\alpha_1$ radiation ($\lambda = 1.5406$~\AA), and data were collected over the range $2\theta = 5^\circ$--150$^\circ$. The sulfur concentration in the sample was analyzed using energy-dispersive X-ray spectroscopy (EDS, Oxford Instruments X-Max) attached to a scanning electron microscope (SEM, Hitachi S-3700N).} This composition did not exhibit superconductivity. To address this, the sulfur-deficient sample underwent a sulfurization process via annealing in the presence of a carefully controlled amount of sulfur. This procedure resulted in the successful synthesis of sulfur-compensated CuCo$_{2}$S$_{4}$, which exhibited bulk superconductivity with a critical temperature, $T_\mathrm{SC}$, of 4.4\,K. Further details on the synthesis protocol and structural, physical, and magnetic characterization are provided in Ref.~\cite{Jin2021}.

In our study, we have used the $\mu$SR technique, which utilizes spin-polarized positive muons to probe the magnetic and superconducting properties of materials. The $\mu$SR measurements were performed on $\sim$1\,g of powder sample, mounted on a \textcolor{black}{haematite masked silver plate (Goodfellow, 99.999\% purity, 0.5~mm thickness) with a hole 6 mm in diameter (28 mm$^{2}$ in area)} using GE varnish to ensure good thermal contact. The sample was cooled down to the lowest temperatures using a sorption cryostat, with the temperature monitored via a calibrated sensor in direct contact with the sample holder to ensure accurate and reliable temperature control. When implanted into a sample, muons rapidly thermalize and precess around the local magnetic field. Muons are unstable particles with a mean lifetime of 2.2~$\mu$s and decay according to the equation $\mu^{+} \rightarrow e^{+} + \overline{\nu}_\mu + \nu_e$. Positrons ($e^{+}$) from this decay are emitted preferentially along the muon spin direction at the moment of decay. The asymmetry in the positron decay count rate, $A_{\mu}(t)$, measured in a specific direction, is directly proportional to the spin polarization $P_\mu(t)$ of the muon ensemble. The asymmetry $A_{\mu}(t)$ is calculated using the relation:
\[
A_{\mu}(t) = \frac{N_{\mathrm{F}}(t) - \alpha N_{\mathrm{B}}(t)}{N_{\mathrm{F}}(t) + \alpha N_{\mathrm{B}}(t)},
\]
where $N_{\mathrm{F}}(t)$ and $N_{\mathrm{B}}(t)$ are the time-dependent positron counts in the forward and backward detectors, respectively, and $\alpha$ is an instrumental calibration constant. The value of $\alpha$ is determined in the normal state under a small transverse magnetic field of approximately 2\,mT. Additional details regarding the detector geometry and experimental methodology are provided in Ref.~\cite{Hillier2022}.

Our $\mu$SR experiments on CuCo$_2$S$_4$ were conducted using the state-of-the-art MuSR spectrometer at the ISIS Pulsed Neutron and Muon Source, Rutherford Appleton Laboratory, Oxford, United Kingdom. Measurements were performed over a temperature range of 0.25 \textcolor{black}{K} to 5.2\,K using a He$^3$ sorption cryostat in both longitudinal- and transverse-field configurations to investigate the superconducting properties of this compound. Zero-field measurements employed an active field compensation system to reduce stray magnetic fields, including the geomagnetic field, down to the milligauss level. For transverse-field (TF) $\mu$SR experiments, an external magnetic field $\mu_{0}H$ was applied above the superconducting transition temperature, and the sample was field-cooled to base temperature to establish a flux-line lattice (FLL). The internal magnetic field distribution, influenced by the magnetic penetration depth $\lambda_{L}$, vortex core radius, and FLL structure, was measured as a function of temperature. Given the reported critical fields $\mu_{0}H_{C1} \sim 10$\,mT and $\mu_{0}H_{C2} \sim 2.46$\,T~\cite{Jin2021}, we selected an applied field of $\mu_{0}H = 30$\,mT to probe the superconducting state. Asymmetry spectra from the $\mu$SR measurements were analyzed using the WiMDA software package~\cite{Pratt2000}.

\section{Results and discussion}
\label{results}

\subsection{Structural Characterization and Physical Properties}

CuCo$_{2}$S$_{4}$ crystallizes in the cubic spinel structure with space group $Fd\overline{3}m$, exhibiting lattice parameters of $a = b = c = 9.468(1)$~\AA~and a unit cell volume of 848.74(3)~\AA$^{3}$. XRD analysis with Rietveld refinement is included in Appendix~\ref{app:XRD}~\cite{Jin2021} . This structure consists of 56 atoms within the unit cell, as illustrated in the schematic diagram in Fig.~\ref{fig1}(a). Copper (Cu) ions occupy the tetrahedral A-sites, while cobalt (Co) ions are situated at the octahedral B-sites. The sulfur (S) ions form a face-centered cubic (fcc) close-packed lattice, providing a robust framework that hosts the metallic ions. The position of Co ions in the octahedral sites plays a key role in determining the electronic band structure and superconducting behavior of CuCo$_{2}$S$_{4}$~\cite{Jin2021}. XRD analysis confirms the presence of a secondary (Co,Cu)S$_{2}$ phase, indicating that the sample is an inhomogeneous two-phase system rather than a homogeneous phase containing magnetic impurities. The majority phase is CuCo$_{2}$S$_{4}$ (85.5\%), while the minority impurity phase is (Co,Cu)S$_{2}$ (14.5\%), which is ferromagnetic with a transition temperature $T_\mathrm{C} = 122$\,K.

The temperature dependence of the magnetic susceptibility $\chi(T)$ of CuCo$_{2}$S$_{4}$ in an applied magnetic field of 1\,mT is shown in Fig.~\ref{fig1}(b). $\chi(T)$ reveals a clear signature of superconductivity below the transition temperature $T_\mathrm{SC} = 4.4$\,K. The observed superconducting volume fraction (calculated as $4\pi \chi$) appears to exceed 100\% due to demagnetization effects (the data are not corrected for this). This is a common artifact in type-II superconductors with non-spherical sample geometry~\cite{Aharoni1998}. The positive value of the susceptibility above $T_\mathrm{SC}$ is attributed to the presence of the ferromagnetic CoS$_2$ impurity in the sample. In addition, the normal-state susceptibility exhibits a weak, nearly temperature-independent behavior, consistent with Pauli paramagnetism~\cite{Jin2021}. The penetration depth is estimated to be $\lambda_L = 241$\,nm using the parameter values of \cite{Jin2021}. Furthermore, the Ginzburg--Landau parameter, $\kappa = \lambda/\xi$, is found to be $\kappa = 20.7$.

The temperature dependence of the heat capacity, \( C_{\mathrm{P}} \), in the range 1.8 to 5\,K in zero applied magnetic field is presented in Fig.~\ref{fig2}(a), where a distinct superconducting transition is evident below 4.4\,K. In the normal state above \( T_{\mathrm{SC}} \), the heat capacity is independent of the applied magnetic field and follows the expression \( C_{\mathrm{P}}(T) = \gamma T + \beta T^{3} \), where \( \gamma \) is the electronic Sommerfeld coefficient and \( \beta T^{3} \) accounts for the phonon contribution. The fit yields \( \gamma = 32.0(2)\,\mathrm{mJ\,mol^{-1}K^{-2}} \) and \( \beta = 0.325(1)\,\mathrm{mJ\,mol^{-1}K^{-4}} \). We next estimate the Debye temperature to calculate the electron-phonon coupling constant ($\lambda_{ep}$) using the McMillan equation. Using the Debye model, the Debye temperature \( \Theta_{\mathrm{D}} \) is determined by $\Theta_{\mathrm{D}} = \left(\frac{12\pi^{4}}{5\beta}nR\right)^{1/3}$, where \( R = 8.314\,\mathrm{J\,mol^{-1}K^{-1}} \) is the gas constant and \( n = 7 \) is the number of atoms per formula unit in CuCo\(_{2}\)S\(_{4}\), giving \( \Theta_{\mathrm{D}} \approx 347(5)\,\mathrm{K} \).

The temperature dependence of the electronic specific heat, \( C_{\mathrm{e}}(T) \), obtained by subtracting the phonon contribution from \( C_{\mathrm{P}}(T) \), is shown in Fig.~\ref{fig2}(b). We analyzed the electronic heat capacity in the superconducting state within the BCS framework~\cite{BCS}. The solid red curve in Fig.~\ref{fig2}(b) represents a fit according to
\[
C_{\mathrm{e}} = C_{\mathrm{BCS}} + \gamma_{\mathrm{ns}} T,
\]
where \( C_{\mathrm{BCS}} \) describes weak-coupling, fully gapped, isotropic $s$-wave BCS superconductivity. The fit yields a superconducting gap of \textcolor{black}{\( \Delta(0) = 0.71(5)\,\mathrm{meV} \)}, consistent with the \( \mu \mathrm{SR} \) measurement discussed later. The term \( \gamma_{\mathrm{ns}} T \) accounts for the nonsuperconducting fraction of the sample.

\subsection{Probing the Symmetry of the Superconducting Energy Gap}

\begin{figure}[t]
\centering
\includegraphics[width=0.9\linewidth]{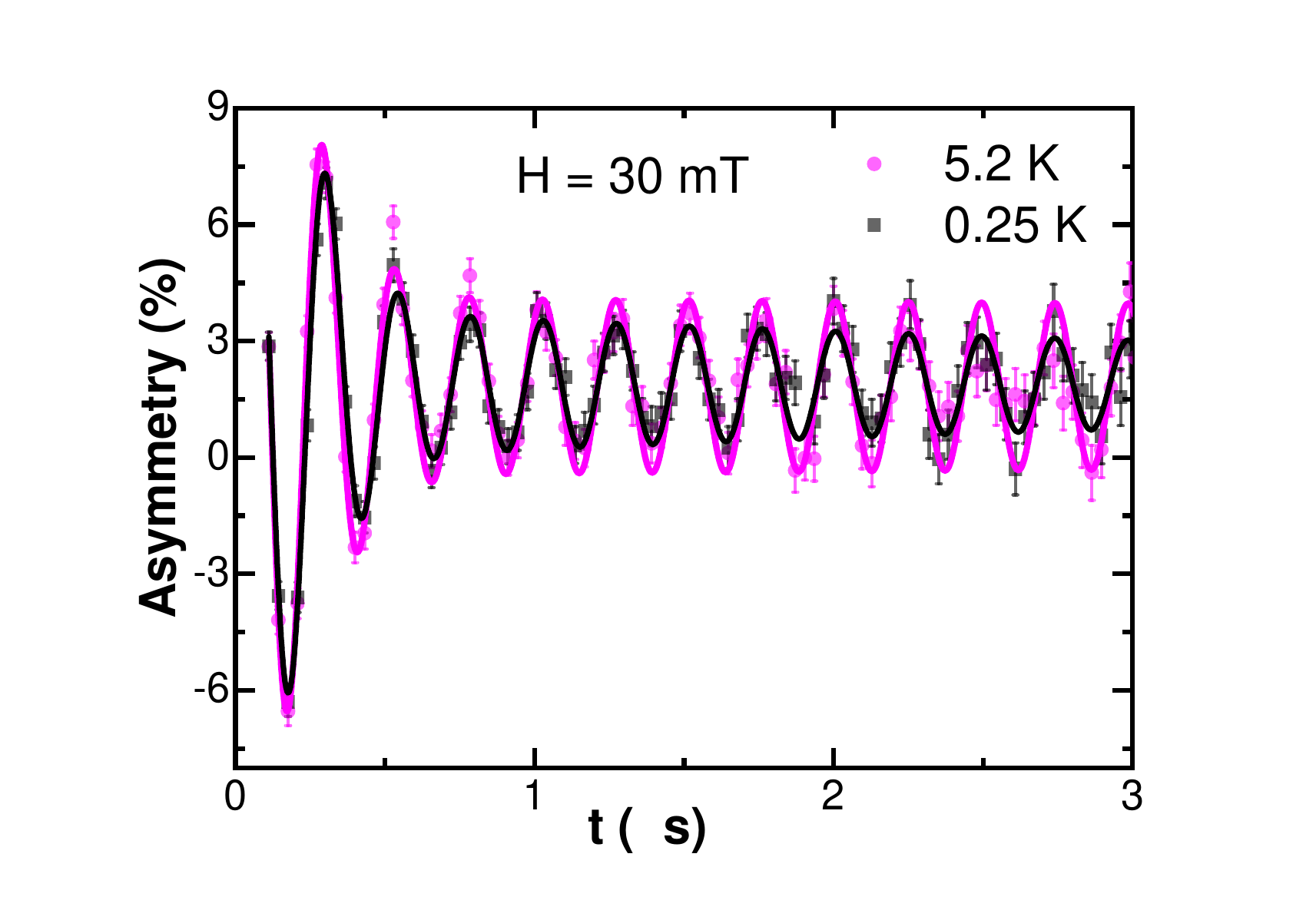}
\caption{Transverse-field $\mu$SR asymmetry spectra for CuCo$_{2}$S$_{4}$ collected at an applied magnetic field of 30\,mT, comparing the field-cooled state at 0.25\,K (below the critical temperature, \(T_\mathrm{SC}\)) and 5.2\,K (above \(T_\mathrm{SC}\)). The solid lines represent fits using Eq.~\ref{TF}.}
\label{fig3}
\end{figure}

To investigate the details of superconducting pairing in CuCo$_{2}$S$_{4}$, we conducted temperature-dependent transverse-field (TF) $\mu$SR measurements. Representative TF-$\mu$SR spectra obtained in both the superconducting and normal states are presented in Fig.~\ref{fig3}. In the normal state, we observed a depolarization of the TF-$\mu$SR signal. In the normal (paramagnetic) state, the relaxation of the TF-$\mu$SR signal arises from two main sources. The first is a weak, temperature-independent depolarization due to the randomly oriented nuclear magnetic moments of Cu and Co. The second, more significant contribution originates from a minor impurity phase of CoS$_{2}$. This impurity exhibits ferromagnetic behavior, with a Curie temperature of \(T_\mathrm{Curie} = 122\)~K and an ordered magnetic moment of \(\mu_{s} = 0.93\,\mu_{B}/\text{Co}\)~\cite{Teruya2017}. Muons stopping in this ferromagnetic impurity phase experience an additional relaxation channel, leading to enhanced depolarization of the TF-$\mu$SR signal in the normal state~\cite{Jin2021}.

In contrast, the superconducting state at 0.25\,K showed enhanced damping in the TF asymmetry spectra. This change indicates the formation of a flux-line lattice characteristic of the superconducting mixed state. \textcolor{black}{X-ray diffraction (XRD) analysis reveals that the synthesized sample consists predominantly of the superconducting CuCo$_{2}$S$_{4}$ phase, with a minor secondary phase identified as ferromagnetic CoS$_{2}$. Rietveld refinement of the diffraction pattern indicates that the impurity phase constitutes approximately 14.5\% of the total sample volume. Since CoS$_{2}$ exhibits ferromagnetic order with a Curie temperature T$_\mathrm{C} \approx$ 122 K, the presence of this phase may introduce additional magnetic relaxation in the $\mu$SR spectra. Therefore, itscontribution has been explicitly considered in the analysis of the TF-$\mu$SR data.} The TF asymmetry spectra were fitted using the following function:
\begin{equation}
\begin{aligned}
A_\mathrm{TF}(t) &= A_\mathrm{sc}\cos\left(\omega_1 t + \varphi\right)\exp\left(-\frac{\sigma_\mathrm{Total}^{2} t^{2}}{2}\right) \\
&\quad + A_\mathrm{im}\exp(-\lambda_\mathrm{TF} t) + A_\mathrm{bg}\cos\left(\omega_2 t + \varphi\right),
\end{aligned}
\label{TF}
\end{equation}
Here, the first term \(\exp\left(-\frac{\sigma_\mathrm{Total}^{2} t^{2}}{2}\right)\) accounts for muon depolarization due to the inhomogeneous fields associated with the vortex state, while \(\exp(-\lambda_\mathrm{TF} t)\) describes relaxation due to magnetic impurities. It is worth mentioning that the damping caused by the impurity is very strong and occurs over a timescale that lies mostly outside the time window of the MuSR instrument at the ISIS facility. This fitting approach is consistent with methodologies employed in other superconducting systems, such as PrOs$_4$Sb$_{12}$~\cite{MacLaughlin2002} and Pr$_{1-x}$Nd$_x$Os$_4$Sb$_{12}$~\cite{MacLaughlin2014}, where superconductivity coexists with magnetic impurities. In our fits, the superconducting asymmetry is \(A_\mathrm{sc} = 73(2)\%\), the impurity component is \(A_\mathrm{im} = 16(1)\%\), and the remaining 11(3)\% corresponds to a non-depolarizing background signal (\(A_\mathrm{bg}\)). \textcolor{black}{These asymmetry values are consistent with the phase fractions obtained from XRD analysis, where the majority phase is CuCo$_2$S$_4$ (85.5\%) and the minority impurity phase is (Co,Cu)S$_2$ (14.5\%).} \textcolor{black}{The background asymmetry fraction (\(A_\mathrm{bg}\)) was determined from the normal-state spectra above T$_\mathrm{SC}$, where the superconducting contribution is absent. Since silver produces  negligible muon spin relaxation, the background component appears as a nearly nonrelaxing oscillatory signal. The extracted background fraction was subsequently fixed or constrained during the fitting of the temperature-dependent TF-$\mu$SR spectra. The resulting value is consistent with the expected fraction of muons stopping in the silver sample holder and remains temperature independent throughout the measurement range.}

The background is temperature independent, \textcolor{black}{determined from the normal-state fit} and is modeled with a separate oscillatory term with frequency \(\omega_2\), while the sample signal precesses with frequency \(\omega_1\), and both share a common initial phase \(\varphi\).

The Gaussian relaxation rate \(\sigma_\mathrm{Total}\) reflects the internal field distribution in type-II superconductors, arising from a combination of the vortex lattice (\(\sigma_\mathrm{sc}\)) and the normal state (\(\sigma_\mathrm{N}\)) of the host material, as described by
\[
\sigma_\mathrm{Total}^{2} = \sigma_\mathrm{sc}^{2} + \sigma_\mathrm{N}^{2}.
\]
While \(\sigma_\mathrm{N}\) remains temperature-independent and is determined from spectra above \(T_\mathrm{SC}\), \(\sigma_\mathrm{Total}\) increases significantly below \(T_\mathrm{SC}\). We note that the normal state relaxation rate $\sigma_{\mathrm{N}}$ includes a contribution from the magnetic impurity phase, and may, in principle, exhibit a weak temperature dependence. To evaluate the validity of assuming $\sigma_{\mathrm{N}}$ to be constant in our analysis, we performed two sets of fits to the TF-$\mu$SR spectra, as detailed in Appendix~\ref{TF-muSR}: one with the impurity-related relaxation rate $\lambda_{\mathrm{TF}}$ held fixed and another allowing it to vary with temperature. As shown in Fig.~\ref{fig:s3}, both approaches reproduce the overall temperature dependence of the total relaxation rate $\sigma_{\mathrm{T}}$, and the variation in $\lambda_{\mathrm{TF}}$ is minor across the temperature range studied, which is expected considering the very ferromagnetic ordering temperature \(T_\mathrm{Curie} = 122\)~K of the impurity phase. In contrast, the superconducting relaxation rate $\sigma_{\mathrm{sc}}$ shows a significant enhancement below \(T_{\mathrm{SC}}\). Therefore, the variation in the impurity-related background relaxation is small compared to the superconductivity-driven changes in $\sigma_{\mathrm{Total}}$, and treating $\sigma_{\mathrm{N}}$ as temperature-independent (in the time window of our analysis) introduces negligible error in our modeling of the superconducting gap.

The normal state contribution to the relaxation rate is significant in this material and likely arises due to a dominant component from the ferromagnetic impurity along with a weaker one from the Cu/Co nuclear moments. In our analysis, we first determined \(\sigma_\mathrm{N} = 3.285(4)\,\mu\mathrm{s}^{-1}\) by averaging data collected in the normal state. With this parameter established, we investigated the temperature dependence of \(\sigma_\mathrm{sc}\), to examine the superconducting response of CuCo$_{2}$S$_{4}$. The observed suppression of \(T_\mathrm{SC}\) under a 30\,mT external magnetic field is consistent with established behavior in the vortex state of superconductors.

To gain deeper insight into the superconducting gap structure of CuCo$_{2}$S$_{4}$, we performed a systematic investigation of the temperature dependence of the magnetic penetration depth and superfluid density using the local London approximation~\cite{Prozorov, Anand2023, Bhattacharyya2024,Zr5Pt3}. The relationship can be expressed as:
\begin{align}
\frac{\sigma_\mathrm{sc}(T)}{\sigma_\mathrm{sc}(0)} &= \frac{\lambda_{L}^{-2}(T,\Delta)}{\lambda_{L}^{-2}(0,\Delta)}, \nonumber \\
&= 1 + \frac{1}{\pi}\int_{0}^{2\pi}\int_{\Delta(T)}^{\infty}\left(\frac{\partial f}{\partial E}\right) \frac{E\,dE\,d\phi}{\sqrt{E^{2}-\Delta^2(T,\phi)}},
\label{Eq2}
\end{align}
where \(f = \left[1+\exp\left(E/k_\mathrm{B}T\right)\right]^{-1}\) is the Fermi distribution function, and \(k_\mathrm{B} = 8.617 \times 10^{-5}\)~eV\,K\(^{-1}\) is Boltzmann's constant. The superconducting gap function \(\Delta(T,\phi)\) is defined as \(\Delta_0 \delta(T/T_\mathrm{SC})\,g(\phi)\), with \(\Delta_0\) representing the zero-temperature gap value. The temperature dependence of the gap is modeled as:
\[
\delta(T/T_\mathrm{SC}) = \tanh\left[1.82\left(1.018\left(T_\mathrm{SC}/T - 1\right)\right)^{0.51}\right].
\]

\textcolor{black}{\(g(\phi)\) contains the angle dependence of the superconducting gap structure, and $\delta(T/T_c)$ the $T$ dependence. $g(\phi)= 1$ for an isotropic $s$-wave gap and two gap ($s+s$), and $g(\phi) = \cos(2\phi)$ for a $d$-wave gap \cite{Pang2015, Annet1990,CeIr3}.}

\begin{figure}[t]
\centering
\includegraphics[width=0.9\linewidth]{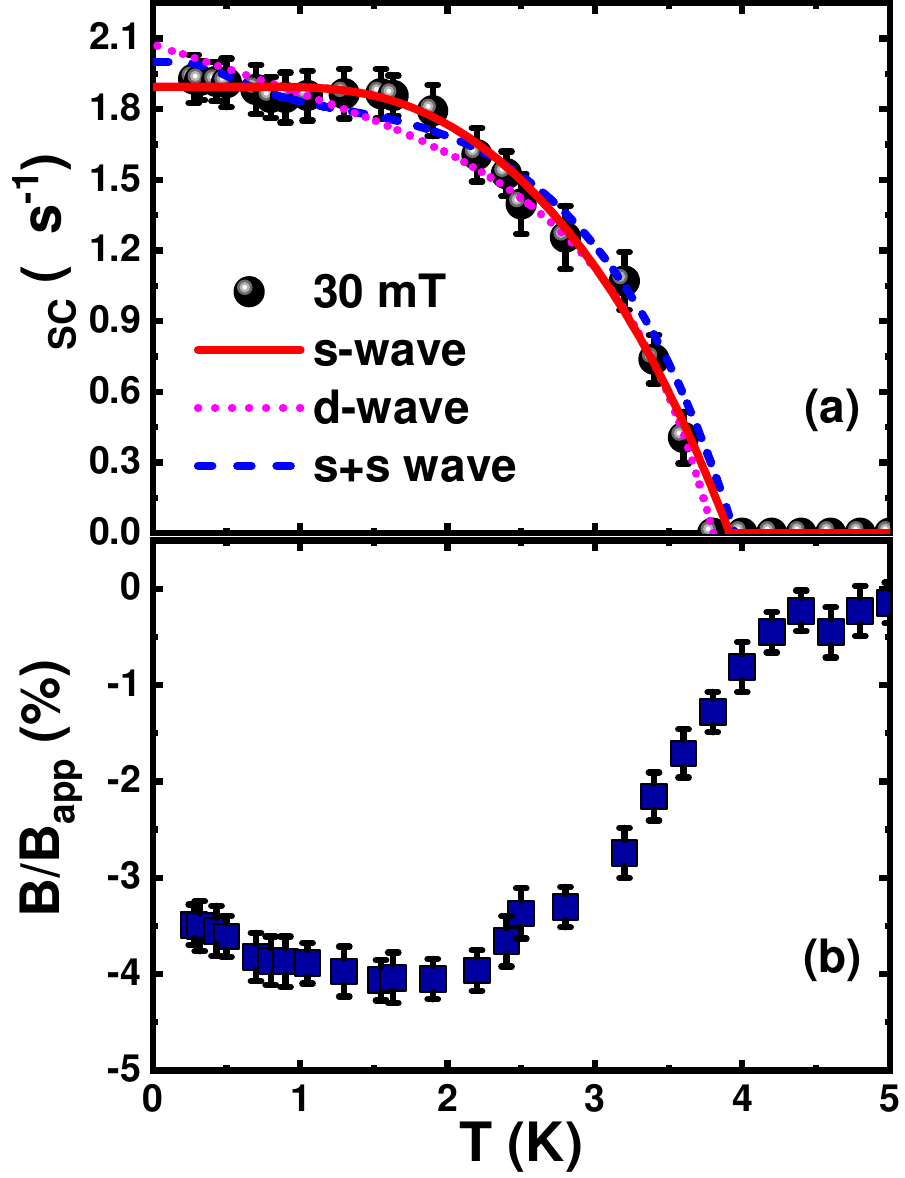}
\caption{Superconducting properties of CuCo$_{2}$S$_{4}$ under an applied magnetic field of 30\,mT. (a) Temperature dependence of the superconducting depolarization rate, \(\sigma_\mathrm{sc}(T)\), with a fit using an \(s\)-wave (red line), $d-$ wave (pink dot) and $s+s$ wave (dashed) gap model. (b) Relative field shift, \(\Delta B = (B_\mathrm{sc} - B_\mathrm{app})/B_\mathrm{app}\), as a function of temperature.}
\label{fig4}
\end{figure}

\textcolor{black}{The solid red curve in Fig.~\ref{fig4}(a) shows the fitting with an isotropic $s$-wave model using the Eq.~(\ref{Eq2}).  We also fitted the $\sigma_\mathrm{SC}(T)$ two-gap ($s+s$)-wave (isotropic) model and $d$-wave model which are also shown in Fig.~\ref{fig4} (a). A closer inspection of the fits by the three models reflects that the single-gap isotropic $s$-wave model describes the $T$ dependence of $\sigma_\mathrm{SC}(T)$ the best. The lower value of goodness of fit parameter $\chi^2$ (see Table~\ref{Table}) also reflects the suitability of isotropic $s$-wave model over the ($s+s$)-wave and $d$-wave models. The values of the energy gaps obtained from the fits with different models are listed in Table~\ref{Table}. We find an energy gap $\Delta(0) = 0.75(1) $~meV from the fit by the isotropic $s$-wave model which yields $2\Delta(0)/k_{\rm B}T_{c} = 3.95(2)$. Complementary NMR measurements estimate this ratio as 4.14, both exceeding the BCS weak-coupling value of 3.53, indicating intermediate coupling strength~\cite{Sugita2000,Wada2001}. These findings classify CuCo$_{2}$S$_{4}$ as a conventional $s$-wave superconductor with intermediate coupling strength.  }

\begin{table}
\caption{Parameters obtained from the fitting of $\sigma_\mathrm{sc} (T)$ for CuCo$_{2}$S$_{4}$ according to Eq.~(\ref{Eq2}) using three models: single isotropic $s$-wave gap model, two isotropic ($s+s$)-wave gap model, and $d$-wave gap model.}
\label{Table} %
\begin{tabular}{|l|c|c|c|}
\hline
Model & $\Delta_{i}(0)$ (meV) & $2\Delta(0)/k_{\rm B}T_{c}$ & $\chi^{2}$ \\ \hline
isotropic $s$-wave & 0.75(1) & 3.95(2) & 1.2 \\ \hline
two gap ($s+s$)-wave & 0.88(2), 0.13(3) & 4.63(7),0.68(5) & 1.8 \\ \hline
 $d$-wave & 1.21 (2) & 6.37(6) & 2.5 \\ \hline
\end{tabular}
\end{table}

\textcolor{black}{For a type-II superconductor in the mixed state forming a hexagonal Abrikosov vortex lattice, the Gaussian muon-spin depolarization rate $\sigma_\mathrm{sc}$ is related to the magnetic penetration depth $\lambda_\mathrm{L}$ through the second moment of the field distribution. In the low-field limit ($b = H/H_{c2} \ll 1$), this relation simplifies to~\cite{Brandt2003}
\[
\sigma_\mathrm{sc}(T) = 0.0431\,\frac{\gamma_\mu \Phi_0}{\lambda_\mathrm{L}^{2}(T)},
\]
where $\gamma_\mu$ is the muon gyromagnetic ratio and $\Phi_0 = 2.07 \times 10^{-15}$~Wb is the magnetic flux quantum.} 

Using this relation, the $s$-wave fit yields a zero-temperature penetration depth of $\lambda_\mathrm{L}(0) = 200(10)$~nm, consistent with the value obtained from magnetization measurements.

For superconductors with a high upper critical field and a hexagonal Abrikosov vortex lattice, the Gaussian muon-spin depolarization rate \(\sigma_\mathrm{sc}\) is inversely proportional to the square of the magnetic penetration depth \(\lambda\), following~\cite{Brandt2003}:
\[
\sigma_\mathrm{sc}(T) = 0.0431\,\frac{\gamma_\mu \Phi_0}{\lambda_\mathrm{L}^{2}(T)},
\]
where \(\Phi_0 = 2.609 \times 10^{-15}\)~Wb is the magnetic flux quantum. The \(s\)-wave fit yields a penetration depth of \(\lambda_\mathrm{L}(0) = 200(10)\)~nm, which is consistent with values obtained from magnetization measurements.

The London model relates the penetration depth \(\lambda_\mathrm{L}(T)\) to the superconducting carrier density and effective mass:
\[
\lambda_\mathrm{L}^{2} = \frac{m^{*}c^{2}}{4\pi n_\mathrm{s}e^{2}},
\]
where \(m^{*} = (1+\lambda_{\mathrm{e-ph}})m_{\mathrm{e}}\) is the effective mass and \(n_\mathrm{s}\) is the superconducting carrier density. Using the McMillan equation~\cite{McMillan},
\textcolor{black}{\[
\lambda_{\mathrm{e-ph}} = \frac{1.04 + \mu^{\ast} \ln(\Theta_{\mathrm{D}} / 1.45 T_\mathrm{SC})}
{(1 - 0.62\,\mu^{\ast}) \ln(\Theta_{\mathrm{D}} / 1.45 T_\mathrm{CS}) - 1.04}
\]
where $\mu^{\ast}$ is a repulsive electron–electron pseudopotential with typical values on the order of 0.13, and $\Theta_\mathrm{D} = 347$~K,} we obtain an electron-phonon coupling constant \(\lambda_\mathrm{e-ph} = 0.592(3)\), a carrier density \(n_\mathrm{s} = 1.12(4) \times 10^{27}\,\mathrm{m}^{-3}\), and an effective mass \(m^{*} = 1.592(3)\,m_{\mathrm{e}}\). Detailed calculations are provided in Refs.~\cite{Chia, Amato}.

The muon precession frequency \(\omega\) is related to the magnetic field \(B\) via \(\omega = \gamma_\mu B\), where the muon gyromagnetic ratio is \(\gamma_\mu/2\pi = 135.5\)~MHz/T. In our analysis, \(\omega_1\) corresponds to the internal field \(B_\mathrm{int}\), modified by diamagnetic screening, while \(\omega_2\) remains nearly temperature independent. Figure~\ref{fig4}(b) shows the temperature dependence of the relative field shift \(\Delta B = (B_\mathrm{int} - B_\mathrm{app})/B_\mathrm{app}\), expressed as a percentage of the applied field. The shift clearly reflects the characteristic diamagnetic response of a type-II superconductor.

\begin{figure}[t]
\centering
\includegraphics[width=0.9\linewidth]{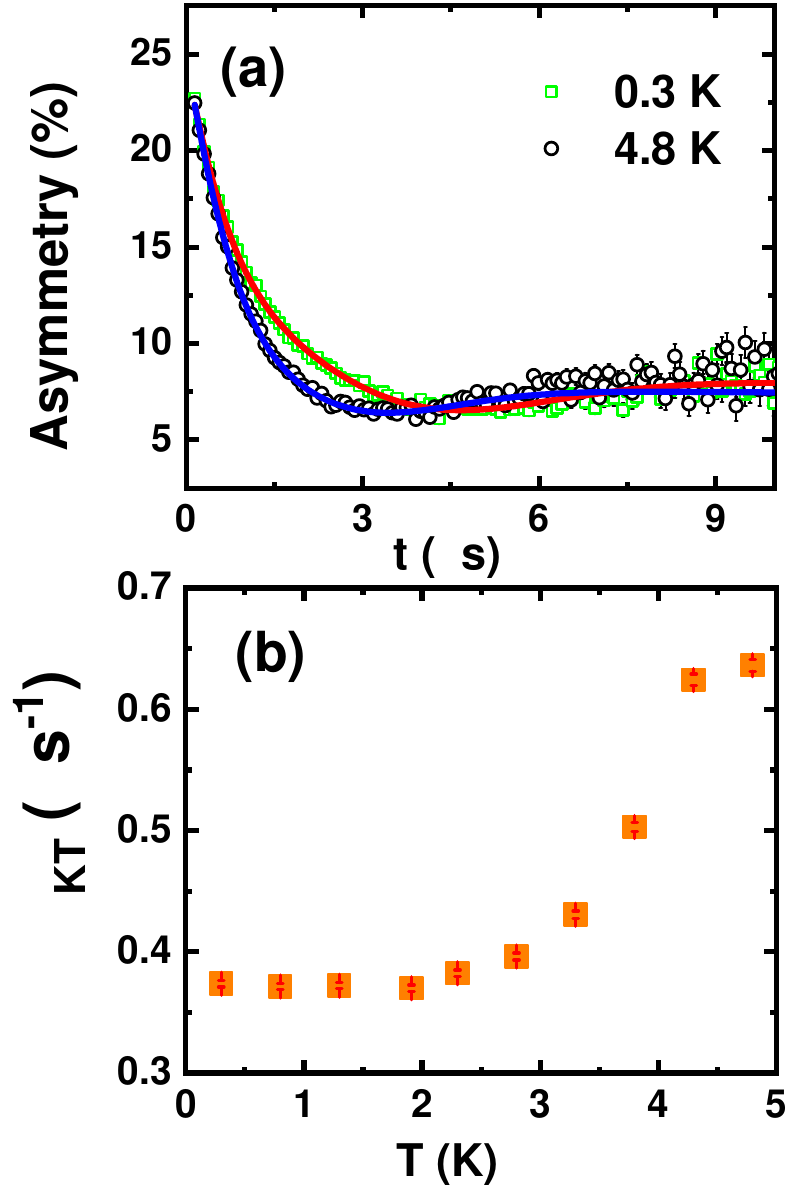}
\caption{(a) ZF asymmetry spectra at 0.3\,K (green squares) and 4.8\,K (black circles). Solid lines are fits to Eq.~\ref{ZF}. (b) Temperature dependence of the relaxation rate associated with the superconducting phase, \(\sigma_\mathrm{KT}\).}
\label{fig5a}
\end{figure}

\begin{figure}[t]
\centering
\includegraphics[width=0.9\linewidth]{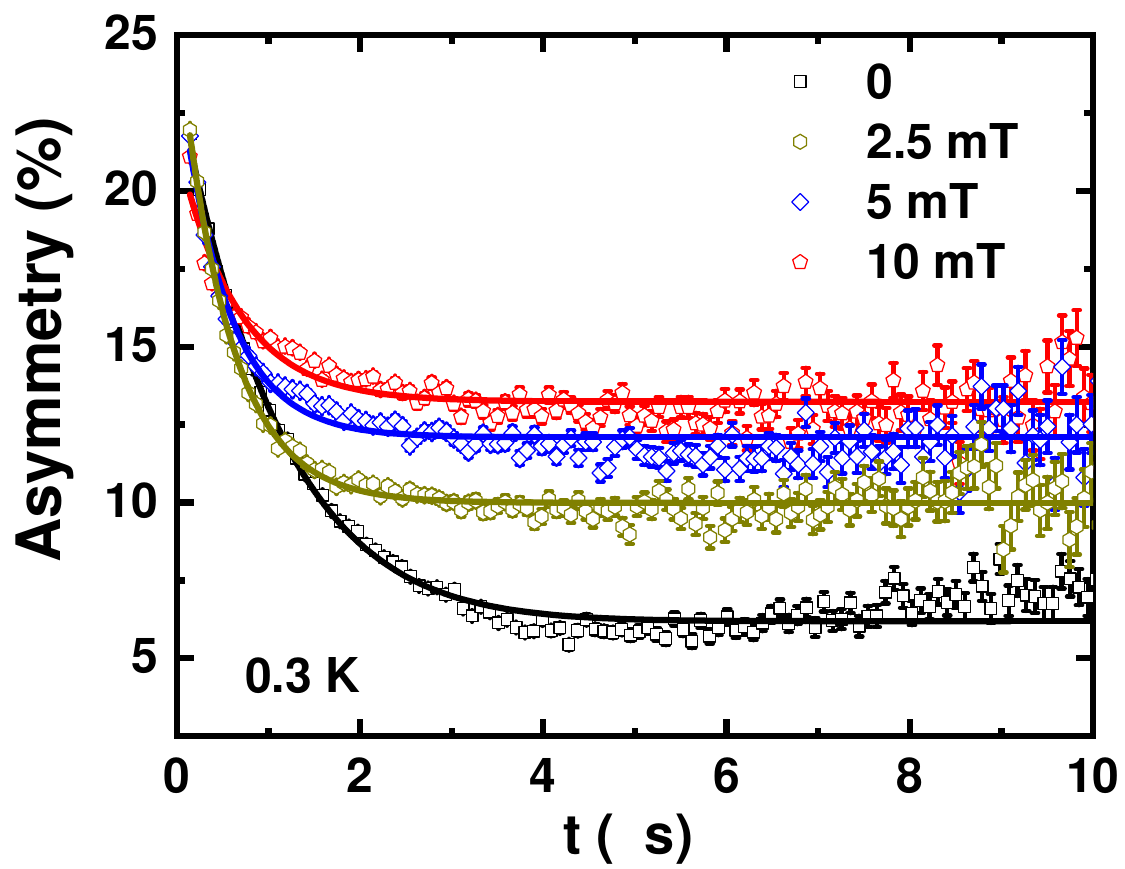}
\caption{LF asymmetry spectra at 0.3\,K under applied fields of 0, 2.5, 5, and 10\,mT. Solid lines are fits using Eq.~\ref{ZF}. The zero-field fit corresponds to that in panel (a) of Fig.~\ref{fig5a}.}
\label{fig5b}
\end{figure}

\subsection{Zero-Field Muon Spin Relaxation}

Representative zero-field (ZF) $\mu$SR asymmetry spectra measured at 0.3\,K (below \(T_\mathrm{SC}\)) and 4.8\,K (above \(T_\mathrm{SC}\)) are shown in Fig.~\ref{fig5a}(a). The spectra contain two main components: a dominant contribution from muons stopping in the sample, and a weak background signal from muons stopping in the silver sample holder. The data were analyzed using a model comprising a damped static Gaussian Kubo-Toyabe (GKT) function~\cite{Hayano1979,CeCo2Ga8}, an exponential relaxation term to model the magnetic impurity, and a constant background:
\begin{equation}
\begin{split}
A_{\text{ZF}}(t) =\, & A_{\text{sc}} \left[ \frac{1}{3} + \frac{2}{3}(1 - \sigma_\text{KT} t) e^{-\sigma_\text{KT} t} \right] \\
& + A_{\text{im}}\,\exp(-\lambda_{\text{im}}t) + A_{\text{bg}},
\end{split}
\label{ZF}
\end{equation}
Here, \(A_\mathrm{sc}\) and \(A_\mathrm{im}\) are the initial asymmetries associated with the superconducting and impurity phases, respectively, and \(A_\mathrm{bg}\) is a temperature-independent background term. Since the impurity phase CoS$_2$ is ferromagnetic below 122\,K, we treat $A_{\text{im}}$ as temperature-independent in our fitting, over the full temperature range studied (0.3--5\,K). The GKT function accounts for the combination of nuclear dipole fields and stray fields from the regions of ferromagnetic impurity. Also as we found from the TF fitting that the relaxation rate associated with the impurity is temperature independent between 0.3\,K and 5\,K, hence we have fixed the value of $\lambda_{\text{im}}$ to 1.447~$\mu$s$^{-1}$ (estimated from 2.8\,K data fit) between 0.3\,K and 5\,K.

The parameter \(\sigma_\text{KT}\) represents the width of the static local field distribution at the muon site, and \(\lambda_\mathrm{im}\) is the relaxation rate due to the impurity phase. The width of the field distribution is given by \(B_\mu = \sigma_\mathrm{KT} / \gamma_\mu\), where \(\gamma_\mu / 2\pi = 135.53\,\text{MHz/T}\) is the muon gyromagnetic ratio. The solid lines in Fig.~\ref{fig5a}(a) represent fits to Eq.~\ref{ZF}. The temperature evolution of \(\sigma_\text{KT}\) is shown in Fig.~\ref{fig5a}(b).

We observe that the damping increases with increasing temperature, in contrast to the behavior reported for materials such as Sc$_5$Co$_4$Si$_{10}$~\cite{Bhattacharyya2022} and CaPd$_2$(Ge,As)$_2$~\cite{Anand2023,Aarti2024}, where an increase in relaxation below \(T_\mathrm{SC}\) is attributed to time-reversal symmetry breaking (TRSB). TRSB has been observed in unconventional superconductors including LaNiC$_2$~\cite{Hillier2009}, Sr$_2$RuO$_4$~\cite{Luke1998}, and UCoGe~\cite{deVisser2009}, which exhibit enhanced low-temperature relaxation.

In contrast, the behavior observed here in CuCo$_2$S$_4$ is consistent with the presence of weak ferromagnetism arising from the CoS$_2$ impurity phase. Since CoS$_2$ is ferromagnetic below 122\,K, it is expected to exhibit a static internal field over the full temperature range studied (0.3--5\,K). Owing to the large Co moment, the transverse relaxation component (2/3 fraction) is extremely rapid and lies outside the ISIS time window, leading to a loss in initial asymmetry. Only the 1/3 longitudinal tail, which is temperature-independent, is visible. This behavior is well known and has been seen in other systems with strong internal fields such as FeSe$_{0.85}$~\cite{Khasanov2008}. Moreover, the observed relaxation trend resembles that in other ferromagnetic systems such as CeCrGe$_3$~\cite{Das2014} and Co/CoO~\cite{vanLierop2003}, where the gaussian width appears to decrease as a function of temperature below $T_\text{Curie}$. The decrease in gaussian width is possibly due to the superconducting phase screening the stray fields generated by the ferromagnetic impurity phase in the sample.

\textcolor{black}{Within the experimental resolution of the present measurements, no clear change in the muon relaxation rate is observed across the superconducting transition temperature. In particular, the ZF-$\mu$SR spectra do not show any additional relaxation or oscillatory signal that would indicate the emergence of spontaneous internal magnetic fields below T$_\mathrm{SC}$. Nevertheless, due to the presence of the ferromagnetic impurity phase and the associated rapid relaxation component, the effective sensitivity of the experiment to weak spontaneous fields is reduced. Consequently, the present ZF-$\mu$SR data should be regarded as inconclusive with respect to the existence of time-reversal symmetry breaking in CuCo$_{2}$S$_{4}$. A definitive test of TRSB in this system would require measurements on phase-pure samples with minimal magnetic background.} 

To distinguish static and dynamic contributions to the relaxation, we performed longitudinal-field (LF) $\mu$SR measurements. Figure~\ref{fig5b} shows LF spectra at 0.3\,K under applied longitudinal fields of 0, 2.5, 5, and 10\,mT. At zero field, a strong relaxation is observed due to static or randomly oriented nuclear dipole fields from the sample and CoS$_2$ impurity. As the field increases, the relaxation is progressively suppressed, indicating partial decoupling of the muon spins from the impurity, where a 10\,mT field is insufficient to fully decouple the muon spins from their local magnetic environment. The inability of a 10\,mT field to fully decouple the relaxation suggests the static fields that are stronger than the applied LF.

\begin{figure}[t]
\centering
\includegraphics[width=0.9\linewidth]{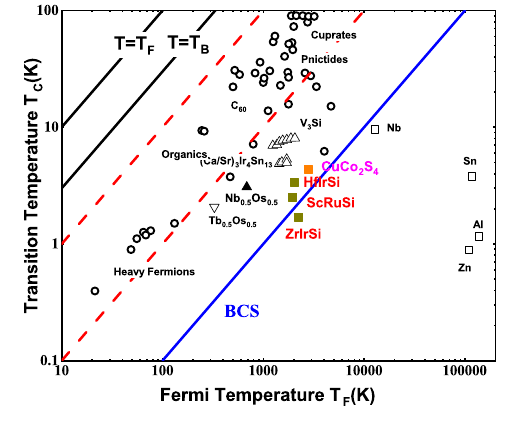}
\caption{Superconducting critical temperature (\(T_\mathrm{SC}\)) versus Fermi temperature (\(T_\mathrm{F}\)) obtained from $\mu$SR measurements in CuCo$_{2}$S$_{4}$. The region between the dashed red lines denotes the range typically associated with unconventional superconductors. The solid blue line marks the position of conventional BCS superconductors for reference. The positions of CuCo$_{2}$S$_{4}$, ScRuSi~\cite{Panda2024}, HfIrSi~\cite{Bhattacharyya2019}, and ZrIrSi~\cite{Panda2019}, located close to the BCS line, indicate that these are conventional superconductors.}
\label{fig6}
\end{figure}

\subsection{Uemura Classification Scheme}

In Fig.~\ref{fig6}, we present a comparison of the critical temperature (\(T_\mathrm{SC}\)) and Fermi temperature (\(T_\mathrm{F}\)) for CuCo$_2$S$_4$, within the framework of the Uemura classification scheme~\cite{Uemura1989,Hillier1997}. The placement of CuCo$_2$S$_4$ outside the unconventional superconductor region confirms its conventional pairing nature. The region between the dashed red lines represents the typical range for unconventional superconductors, while a solid blue line denotes the position of conventional BCS superconductors.

The Fermi temperature is calculated as
\[
T_\mathrm{F} = \frac{\hbar^{2}(3\pi^{2})^{2/3}n_\mathrm{s}^{2/3}}{2k_\mathrm{B}m^{*}},
\]
based on the carrier density \(n_\mathrm{s}\) and effective mass \(m^{*}\), both derived from $\mu$SR measurements of the superconducting penetration depth.

In the Uemura plot, unconventional superconductors typically exhibit \(T_\mathrm{SC}/T_\mathrm{F}\) ratios in the range of 0.001--0.1, whereas conventional Bardeen-Cooper-Schrieffer (BCS) superconductors appear further to the right, with smaller ratios (\(T_\mathrm{SC} / T_\mathrm{F} \leq 1/1000\)). The position of CuCo$_2$S$_4$ in the Uemura plot confirms its conventional BCS-like superfluid density. For CuCo$_2$S$_4$, with \(T_\mathrm{SC} = 4.4\,\text{K}\) and \(T_\mathrm{F} = 2802\,\text{K}\), the resulting ratio
\[
\frac{T_\mathrm{SC}}{T_\mathrm{F}} = \frac{4.4}{2802} = 0.00157
\]
places it near the region associated with conventional superconductors in the Uemura classification.

\section{Summary}
\textcolor{black}{In summary, we have carried out a comprehensive investigation of the superconducting properties of the thiospinel compound CuCo$_2$S$_4$ using muon spin rotation and relaxation ($\mu$SR) measurements, complemented by magnetization and heat-capacity experiments. The temperature dependence of the superconducting depolarization rate obtained from transverse-field $\mu$SR measurements indicates a fully gapped superconducting order parameter. The extracted superconducting gap ratio, $2\Delta(0)/(k_B T_\mathrm{SC}) = 3.95(2)$, exceeds the BCS weak-coupling value of 3.53 and is consistent with previous NMR results, placing CuCo$_2$S$_4$ in the intermediate electron--phonon coupling regime. Zero-field $\mu$SR measurements were performed to probe possible time-reversal symmetry breaking (TRSB) in the superconducting state. Within the experimental resolution, no additional spontaneous internal magnetic fields are observed below $T_\mathrm{SC}$. However, due to the presence of a ferromagnetic CoS$_2$ impurity phase and the associated fast-relaxing signal component, the sensitivity of the present measurements to weak spontaneous fields is reduced. Consequently, although no evidence for TRSB is detected, the possibility of weak TRSB fields cannot be definitively excluded. We also determined key superconducting parameters from the $\mu$SR and thermodynamic measurements, providing a comprehensive characterization of the superconducting state in CuCo$_2$S$_4$. Although the presence of a minor CoS$_2$ impurity phase introduces additional complexity in the interpretation of the $\mu$SR spectra, its contribution has been carefully accounted for in the analysis, and the principal conclusions regarding the fully gapped superconducting state remain robust. Future investigations on phase-pure samples or single crystals would be highly desirable to further clarify the superconducting gap symmetry and to enable a more sensitive test for possible time-reversal symmetry breaking in this system. Nevertheless, the present results provide the first microscopic evidence that CuCo$_2$S$_4$ hosts a fully gapped superconducting state with intermediate coupling strength, consistent with conventional $s$-wave superconductivity in this cobalt-based thiospinel.}

\section*{Acknowledgements}

A.B. acknowledges support from the Science \& Engineering Research Board under the CRG Research Grant (Grant No.~CRG/2022/008528) and the CRS Project at UGC-DAE CSR (Grant No.~CRS/2021-22/03/549). D.T.A. is grateful for financial support from EPSRC UK (Grant Reference: EP/W00562X/1) and the Chinese Academy of Sciences for a PIFI fellowship. G.H.C. acknowledges support from the National Key Research and Development Program of China (Grant Nos.~2023YFA1406101 and 2022YFA1403202).

\appendix

\section{X-Ray Diffraction and Magnetization}\label{app:XRD}

\begin{figure}[h!]
    \centering
    \includegraphics[width=0.7\linewidth]{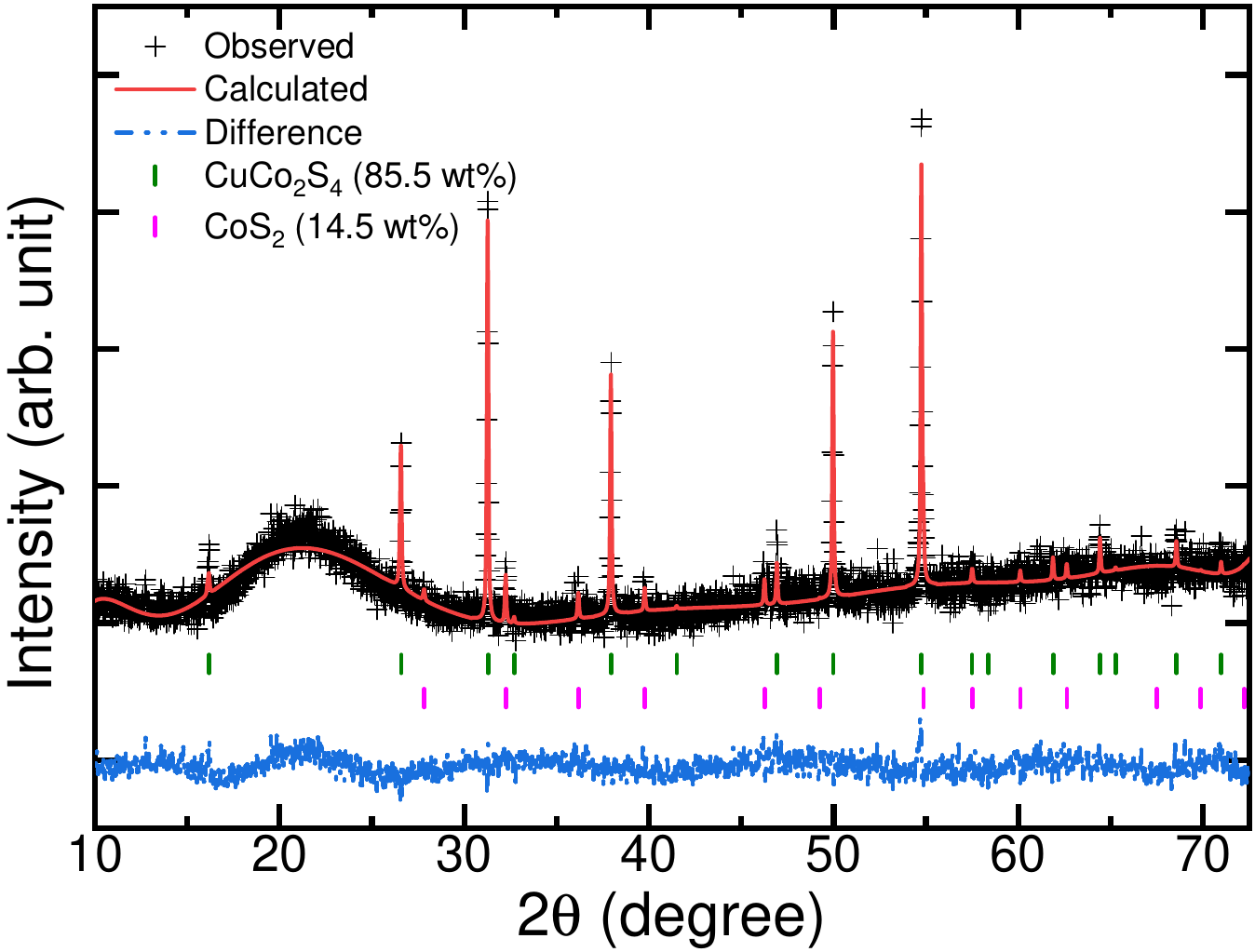}
    \caption{Powder X-ray diffraction pattern and two-phase Rietveld refinement of the CuCo$_2$S$_4$ sample, indicating the presence of a (Co,Cu)S$_2$ secondary phase.}
    \label{fig:s1}
\end{figure}

\begin{figure}[h]
    \centering
    \includegraphics[width=0.8\linewidth]{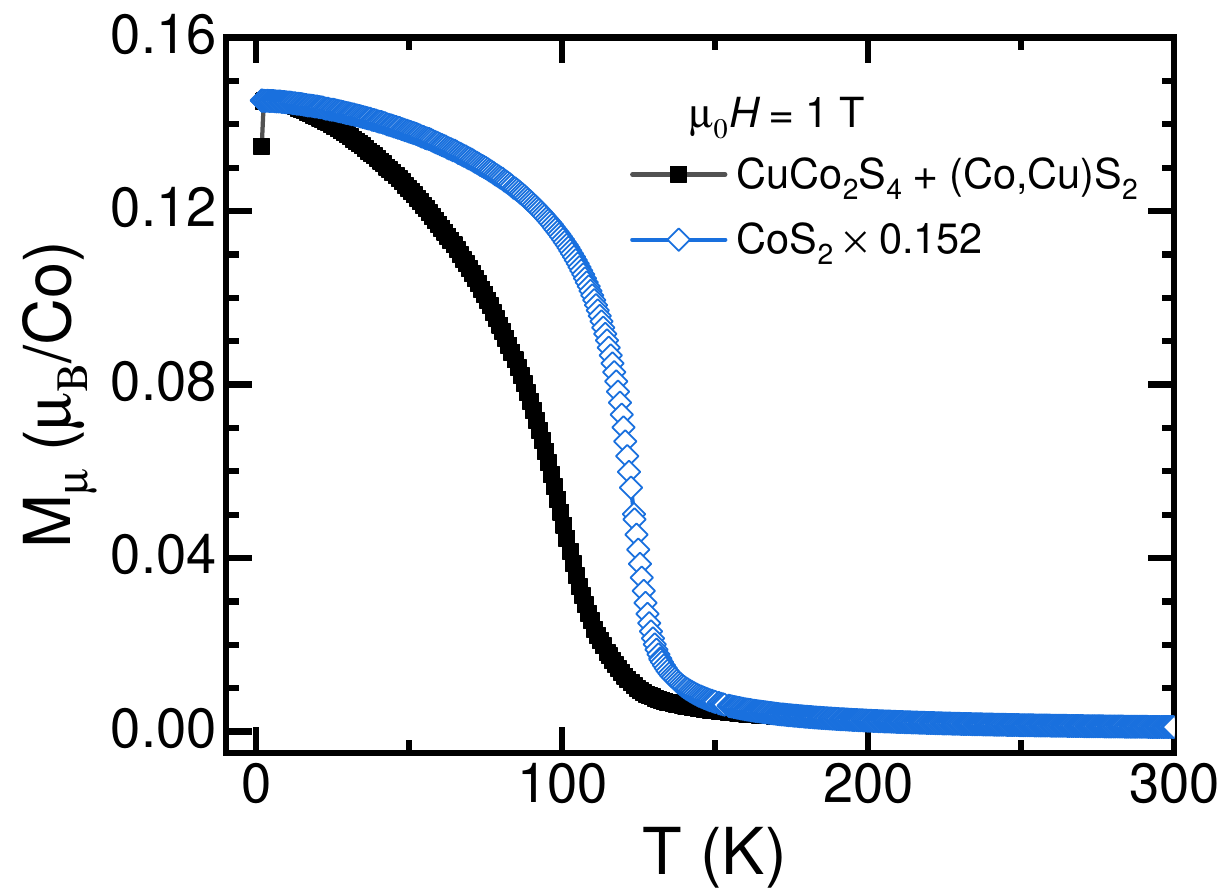}
    \caption{Temperature dependence of magnetization (in $\mu_{B}$/Co) for the CuCo$_{2}$S$_{4}$ sample containing a ferromagnetic secondary phase of (Co,Cu)S$_{2}$. The blue data points were measured using a pure CoS$_{2}$ sample (scaled by a factor of 0.152) for comparison. The fraction of (Co,Cu)S$_{2}$ in the CuCo$_{2}$S$_{4}$ sample is estimated to be approximately 14.5\%, consistent with the XRD analysis in Fig.~\ref{fig:s1}.}
    \label{fig:s2}
\end{figure}

Figure~\ref{fig:s1} shows the powder X-ray diffraction pattern of CuCo$_{2}$S$_{4}$ collected at room temperature, along with the Rietveld refinement profile. The refinement confirms the cubic spinel structure of CuCo$_{2}$S$_{4}$. XRD analysis confirms the presence of a secondary (Co,Cu)S$_{2}$ phase, indicating that the sample is an inhomogeneous two-phase system, rather than a homogeneous phase containing magnetic impurities. Based on a two-phase Rietveld analysis ($R_{wp} = 6.34\%$, $\chi^{2} = 1.01$), the impurity fraction is estimated to be 14.5~wt.\%. This result is consistent with our previous report~\cite{Jin2021}.

In addition, following the method described in Ref.~\cite{Jin2021}, we estimated the amount of ferromagnetic CoS$_{2}$ impurity by comparing magnetization data with that of a single-phase CoS$_{2}$ sample, as shown in Fig.~\ref{fig:s2}. The analysis yields an impurity fraction of approximately 15\%, consistent with the XRD result. The CoS$_{2}$ phase is likely Cu-doped to a small extent.

\section{TF-$\mu$SR}\label{TF-muSR}

\begin{figure}[h]
    \centering
    \includegraphics[width=0.8\linewidth]{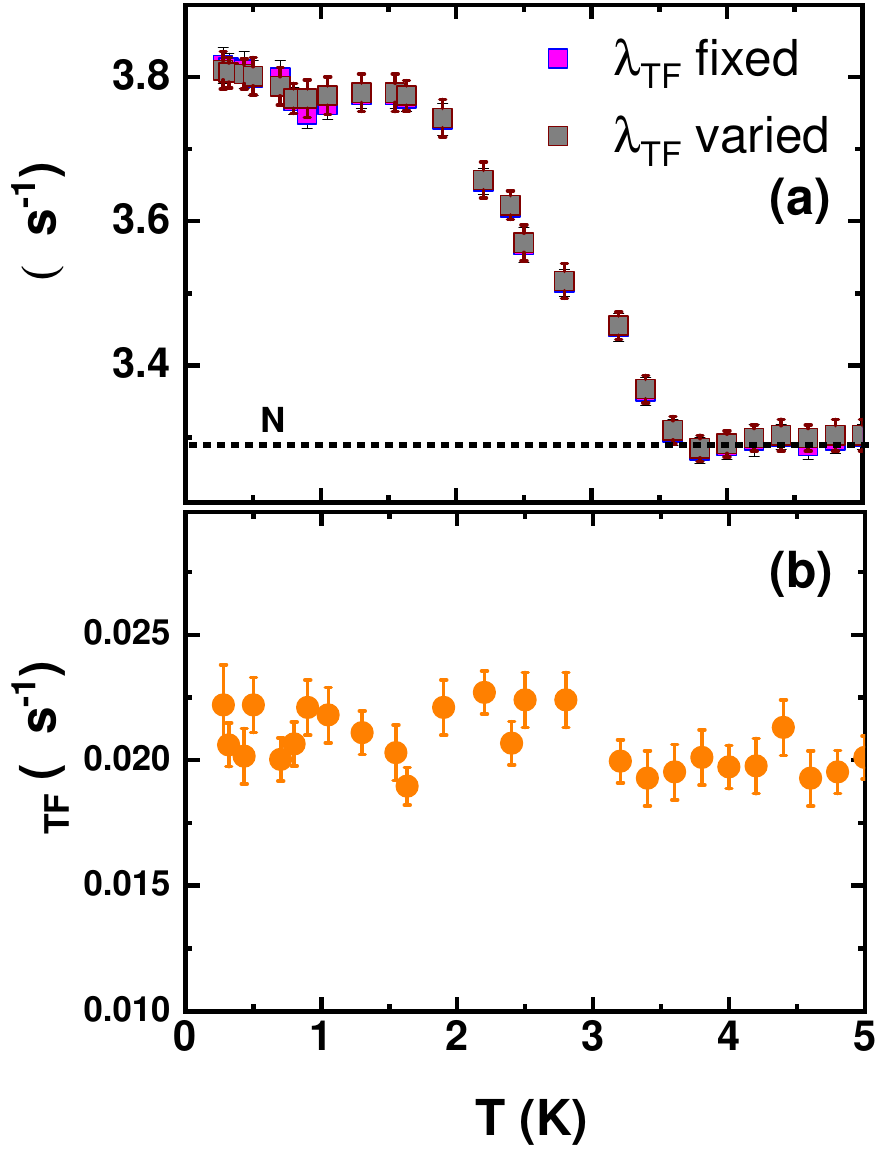}
    \caption{(a) Total muon spin relaxation rate \(\sigma_\mathrm{T}\) as a function of temperature, obtained by fitting with the \textcolor{black}{impurity relaxation rate}\(\lambda_\mathrm{TF}\) either fixed (pink squares) or varied (gray squares). Both superconducting and normal-state contributions are included. (b) The \(\lambda_\mathrm{TF}\) values as a function of temperature for the variable fit case.}
    \label{fig:s3}
\end{figure}

Figure~\ref{fig:s3} shows the temperature dependence of the total muon spin relaxation rate, \(\sigma_\mathrm{T}\), obtained from transverse-field $\mu$SR measurements. The data were analyzed by fitting the asymmetry spectra using two different approaches: one in which \(\lambda_\mathrm{TF}\) was held fixed (pink squares), and another in which \(\lambda_\mathrm{TF}\) was treated as a temperature-dependent free parameter (gray squares).

Panel (a) displays \(\sigma_\mathrm{T}(T)\) for both cases. Both fitting approaches qualitatively reproduce the overall trend. In further calculation we have used the constant \(\lambda_\mathrm{TF}\) fitting. Panel (b) presents the temperature dependence of \(\lambda_\mathrm{TF}\) extracted from the variable fit.

\bibliographystyle{elsarticle-num}  
\bibliography{ref2}

\end{document}